\documentclass[10pt, aps]{revtex4}
\usepackage{amssymb}
\usepackage{latexsym}
\usepackage{epsfig}
\usepackage{float}
\usepackage{graphicx, epsfig, color}
\usepackage{graphicx}
\usepackage{subfigure}
\usepackage{amsmath}
\usepackage{hyperref}

\begin{document}
\title{\textbf{Medium effect on anisotropic surface tension of magnetized quark matter}}
\author{Yu-Ying He, Xin-Jian Wen\footnote{wenxj@sxu.edu.cn} }
\affiliation{ Institute of Theoretical Physics, State Key Laboratory
of Quantum Optics and Quantum Optics Devices,Shanxi University,
Taiyuan, Shanxi 030006, China}

\begin{abstract}

The thermodynamics of finite size quark matter in the quasiparticle
model is self-consistently constructed by an effective bag function,
which presents the medium effect to the confinement. We obtained
completely analytic surface tension in the strong magnetic field
with the multiple reflection expansion. The anisotropic structure is
demonstrated by the splitting of the longitudinal and transverse
surface tensions. The anisotropy of the surface tension could be
enhanced by an increase of the magnetic field. The analytical
surface tension is modified by an additional term related to the bag
function. For strong enough magnetic fields, the increase of the
longitudinal surface tension is proportional to the magnetic field.
On the contrary, the transverse component vanishes due to all quarks
locating in the lowest landau level.
\end{abstract}



\maketitle

\section{Introduction}
 Over the years, many works have been dedicated to the
effects of magnetic fields on the quantum chromodynamics (QCD) phase
transition \cite{Bali:2011qj,Kharzeev:2013jha} and the equation of
state in quark (neutron) stars
\cite{Pal:1998jb,Chakrabarty:1995uja,Isayev:2011zz,Menezes:2015fla,Ferrer:2019xlr,Chaudhuri:2022oru}.
The magnetic field modifies the microscopic properties of quark
matter with the corresponding macroscopic implication in compact
stars
\cite{Chakrabarty:1996te,Broderick:2000pe,PerezMartinez:2005av,Deb:2021ftm}.
It is well known that in the presence of a magnetic field, the
anisotropic effect becomes significant and non-negligible in strong
magnetic field, owing to the breaking of the spational rotational
symmetry
\cite{Canuto:1968apg,Chaichian:1999gd,Martinez:2003dz,Felipe:2002wt,Ferrer:2010wz,Karmakar:2019tdp}.
To reflect the anisotropic structure with a rapid longitudinal
expansion of QGP created in HICs, the anisotropic Coulomb potential
can be produced through an angle-averaged screening mass
\cite{Dong:2021gnb}. Many theoretical works presented the analytic
expression for anisotropic pressures under certain approximation
\cite{Menezes:2015fla,Ferrer:2019xlr,Ferrer:2010wz,Felipe:2007vb}.
Ferrer et al. showed the anisotropic pressure and estimated the
threshold field that separates the isotropic and anisotropic regimes
\cite{Ferrer:2010wz}. Later, Isayev and Yang confirmed the splitting
of the longitudinal and transverse pressure in their articles
\cite{Isayev:2011zz,Isayev:2011ug}. The anisotropic pressure will
affect the determination of the compressibility. The compressibility
could manifest the anisotropic structure due to the breaking of the
rotation symmetry. The discontinuity of longitudinal compressibility
with the chemical potential and the temperature captures the
signature of a first-order chiral phase transition
\cite{Yang:2021rdo}. With increasing temperature, the appearance of
the longitudinal instability prevents the formation of a fully
spin-polarized state in neutron matter and only the states with
moderate spin polarization are accessible \cite{Isayev:2011zz}.
Recently, Lugones et. al. pushed the investigation of the surface
tension in longitudinal and transverse components with respect to
the magnetic field in the bag model
\cite{Lugones:2018qgu,Lugones:2016ytl,Grunfeld:2020gnv}. In this
paper, our aim is to investigate the relevant anisotropic surface
tension reflecting the breaking of the {\it O}(3) rotational
symmetry in the deconfinement process.

In principle, the surface tension together with the QCD phase
diagram should be investigated in the underlying fundamental theory,
lattice QCD (LQCD). However, current LQCD methods are not sufficient
to determine the matter structure at larger chemical potentials. The
only available methods at relatively low energy are effective
models. In literature, the phenomenological models overcome the
difficulty of the QCD theory at finite chemical potentials. In order
to interpret the chiral phase transition and dynamical symmetry
breaking, the Nambu-Jona-Lasinio (NJL) model is widely used in the
QCD-like investigation. The NJL model has proved to be very
successful in the description of the spontaneous breakdown of chiral
symmetry exhibited by the true (nonperturbative) QCD vacuum. It
explains very well the spectrum of the low lying mesons which is
intimately connected with chiral symmetry as well as many other low
energy phenomena of strong interaction. The quark quasiparticle
model, as an extended bag model, has been developed in studying the
bulk properties of the dense quark matter at finite density and
temperature. To describe the strong interaction effects in terms of
effective fugacities, the effective fugacity quasiparticle model is
proposed by Chandra and Ravisankar
\cite{Chandra:2011en,Chandra:2007ca}. The advantage of the
quasiparticle is the successful description of the confinement
mechanism by the density- and/or temperature-dependent bag function,
via which the first-order deconfining phase transition was
constructed and the critical end point was determined
\cite{Srivastava:2010xa}. The aim of this work is to investigate the
anisotropy of surface tension modified by the medium effect in
strong magnetic fields. We also hope that the surface tension is
helpful to investigate the deconfinement transition, since the
surface tension is relevant for bubble nucleation of quark matter in
supernovae \cite{Jimenez:2017fax}. It can play an important role in
the hadron-quark phase transition in the presence of a magnetic
field when the anisotropic approach is followed as it was done
recently \cite{Ferrer:2020tlz}.

This paper is organized as follows. In Section \ref{sec:model}, we
present the self-consistent thermodynamics of the magnetized quark
matter in the quasiparticle model. The medium effect is included by
introducing the effective bag function. The surface tension is
modified by an additional term dependent on the bag function. In
Section \ref{sec:result}, the numerical results for the confinement
bag function and surface tension are shown in the strong magnetic
field. The detailed discussions are focused on the anisotropy of the
surface tension. The last section is a short summary.

\section{Thermodynamics of quasiparticle model in strong magnetic fields}\label{sec:model}

The main purpose of this paper is to study the properties of the
deconfined quark matter in strong magnetic fields. The nonzero quark
masses are explored and the exact chiral symmetry are broken. The
effective quasi-particle mass should be introduced to include the
interaction effect in the quasiparticle approach. The total energy
in the ensemble of quasiparticle as a free and degenerate Fermion
gas can be written as
\begin{eqnarray} H_\mathrm{eff}=\sum_{i=1}^d\sum_p
\sqrt{p^2+{m_i^*}^2}\hat{a}^\dagger_{i,p} \hat{a}_{i,p}+B^*(\mu)V
\end{eqnarray}
where $d$ denotes the degree of the degeneracy including the flavor,
color and spin. The chemical potential-dependent bag function $B^*$
denotes the energy difference between the physical vacuum and the
perturbative vacuum, which is necessary to ensure thermodynamic
consistency.

For the medium dependence of the quark quasiparticle model, the
effective quark mass $m_i^*$ is derived at the zero momentum limit
of the dispersion relation following from the effective quark
propagator by resuming one-loop self energy diagrams in the hard
dense loop (HDL) approximation \cite{Schertler:1996tq}. The
in-medium effective mass of quarks can thus be expressed as
\cite{Schertler:1996tq,Schertler:1997vv,Peshier:1999ww,Bannur:2007tk,Wen:2009zza}
\begin{equation}\label{mass}
m_i^*(\mu_i)=\frac{m_i}{2}+\sqrt{\frac{m_i^2}{4}+\frac{g^2\mu_i^2}{6\pi^2}},
\ \ (i=u,d,s),
\end{equation} where $m_i$ is the current mass of corresponding
quarks and the constant $g$ is related to the strong interaction
constant $\alpha_s$ by the equation $g=\sqrt{4\pi \alpha_s}$. The
quasi-particle idea can be recalled backward to the work by Fowler
et.al.\cite{Fowler:1981rp} that the particle mass may change with
the environment parameters. Following the original ideas, the quark
mass density dependent model is studied by Chakrabarty
et.al.\cite{Chakrabarty:1989bq}. Accordingly, as a phenomenological
method, our quasiparticle model has a similar treatment. They are
apparently different in approach but equally satisfactory in result.
The in-medium screening mass in Eq.(\ref{mass}) is merely a model
assumption on the quasiparticle mass in the present treatment, and
can not be justified field-theoretically. The effective mass depends
on the leading term of quark self-energy in the hard dense loop
approximation. So the expression is only valid for a large chemical
potential.

In order to investigate the finite size effect, we apply the
multiple reflection expansion (MRE). It is originally proposed by
Balian and Bloch in the distribution of eigenvalues of the wave
equation inside the volume bounded by s closed
surface\cite{Balian:1970fw}. The eigenvalue density is smoothed by
the asymptotic expansion to eliminate its fluctuation part due to
the discrete eigenvalues on boundary condition. Later the MRE is
developed in finite size quark matter by Madsen
\cite{Madsen:1993ka},Farhi and Jaffe \cite{Farhi:1984qu} and Berger
and Jaffe \cite{Berger:1986ps}. In the MRE framework, the finite-
size effects are considered in the modified density of state and the
thermodynamic potential density is
\begin{eqnarray}\label{eq:omega}
\Omega_i &=&
  -d_i\frac{2 T}{(2\pi)^3}\int
    \left\{
 \ln\left[1+e^{-(\sqrt{p^2+m_i^2}-\mu_i)/T}\right]
    \right. \nonumber\\
&&  \left. \phantom{-T}
+\ln\left[1+e^{-(\sqrt{p^2+m_i^2}+\mu_i)/T}\right]
    \right\} \rho_{\text{MRE}} \mbox{d}^3p ,
\end{eqnarray}
where $T$ is the system temperature and $d_i=3$ is the color
degeneracy factor for $i$-type quarks. The density of state for a
spherical system is modified by the factor in the multi-expansion
approach by \cite{Balian1970AP60}
\begin{equation} \label{tmd}
\rho_{\text{MRE}}(p,m_i,R)
  =d_i\left[
     1
      +\frac{2\pi^2}{p} \frac{S}{V} f_{\mathrm{S}}\left(x_i\right)
    \right].
\end{equation}
Here $x_i\equiv m_i^*/p$ is the ratio of the quark mass $m_i$ over
the kinetic moment $p$. The dimensionless function
$f_{\mathrm{S}}(x_i)= -\frac{1}{4\pi^2}\mbox{arctan}(x_i)$ would
play an important role in the modification of the density of state
\cite{Berger:1986ps,Farhi1984PRD30}. In conventional form of MRE,
the expansion includes the boundary surface and its curvature the
density of states for a volume of arbitrary shape. In our work, the
curvature term has no influence on our subject and is omitted. For
the extremely relativistic particle with $m\ll p$, the density
modification is not modified significantly \cite{Berger:1986ps}.
Generally, the modification of the density of state in the MRE
framework constrain the low limit on the infrared cutoff due to the
fact that $\rho_\text{MRE}$ becomes negative at small momenta
\cite{Kiriyama:2005eh}.

In the integrations of the thermodynamic potential, the following
replacement should be applied
\begin{eqnarray} \int \frac{d^3p}{(2\pi)^3} \rightarrow
\frac{|q_iB_m|}{2\pi}\sum_\nu \sum_{s=\pm1}\int_0^\infty
\frac{dp_z}{2\pi},
\end{eqnarray}where the magnetic field strengths the degeneracy factor $|qB|$ together with the
dimensional reduction. At zero temperature and strong magnetic
fields, Eq.(\ref{eq:omega}) is simplified as
\begin{eqnarray}\label{thero-int}
\Omega_i &=& \frac{d_i |q_iB_m|}{2\pi^2}
\sum_{\nu=0}^{\nu_i^\mathrm{max}} (2-\delta_{\nu 0})
  \int_{\Lambda_\mathrm{IR}}^{p_F}
    \left(
E_i-\mu_i
    \right)
\rho_\text{MRE}  \mbox{d}p_z\, ,
\end{eqnarray}
where the single particle energy eigenvalue
$E_i=\sqrt{{m_i^*}^2+p_z^2+ 2 \nu_i |q_i B|}$ sensitively depends on
the magnetic fields. The infrared cutoff of the momentum
${\Lambda_\mathrm{IR}}$ is required to obtain the non-negative
density of state. At zero temperature, the upper limit
$\nu_i^\mathrm{max}$ of the summation index $\nu_i$ can be
understood from the positive value requirement on Fermi momentum and
is defined by
\begin{eqnarray} \nu_i^\mathrm{max}=\frac{\mu_i^2-{m_i^*}^2}{2 |q_i B_m|}.
\end{eqnarray}

\subsection{The effective bag function}
In the framework of the quasiparticle model, the total thermodynamic
potential with effective mass $m_i^*(\mu_i)$ should be
self-consistently written as
\begin{eqnarray}\Omega=\sum_i [\Omega_i(\mu_i,m_i^*(\mu_i) )+B_i^*(\mu_i)]+B_0,
\end{eqnarray}
where the additional term $B^*_i(\mu_i)$ is the medium dependent
quantity to be determined. The vacuum energy density $B_0$ is
medium-independent. It has been interpreted as a background field,
zero point energy density, or bag pressure \cite{Bannur:2006hp}. In
the standard statistical mechanics, the Hamiltonian or the
thermodynamic potential depends on the temperature and the chemical
potential related to the conserved charges. If the thermodynamic
potential depends on the state variable implicitly via
phenomenological parameters $m_i^*(T,\mu_i)$, the corresponding
stationarity condition should be required as
\cite{Gorenstein:1995vm}
\begin{eqnarray} \left.\frac{\partial \Omega}{\partial
m_i^*}\right |_{T,\mu_i}=0.
\end{eqnarray}
which has been widely employed in the quasiparticel model at finite
temperature and density
\cite{Peshier:1999ww,Schneider:2001nf,Bannur:2006js,Giacosa:2010vz}.
The condition respects the chiral symmetry restoration in the plasma
\cite{Peshier:1998ei}. At zero temperature, we get the bag function
through the integral
\begin{eqnarray}\label{Bexp}
B_i^*(\mu_i)&=&- \frac{d_i |q_iB_m|}{2\pi^2}
\sum_{\nu=0}^{\nu_i^\mathrm{max}} (2-\delta_{\nu 0})
\int^{\mu_i}_{m_i^*} \int_{\Lambda_\mathrm{IR}}^{p_F}
(\frac{m_i^*}{E_i}\rho_\text{MRE}-\frac{3}{2R}
\frac{E_i-\mu_i}{{E_i}^2} )\frac{\mathrm{d}
m^*_i}{\mathrm{d}\mu_i}\mathrm{d}p_z \mathrm{d}\mu_i,
\end{eqnarray}
where the lower limit means the allowed chemical potential.
Specially, the equality $\mu_i=m_i^*$ would lead to the vanishing
Fermi momentum and the zero bag function.

According to the geometric dependence, namely, the power of the
radius dependence,  the thermodynamic potential density can be
considered as
\begin{eqnarray}
\Omega_i=\underbrace{\Omega_{V,i}+B_{V,i}}_\text{volume}
+\underbrace{ \Omega_{S,i}+B_{S,i}}_\text{surface}. \label{BmuR}
\end{eqnarray}
The surface terms proportional to the $1/R$ are
\begin{eqnarray}\Omega_{S,i}&=&-\frac{d_i |q_iB_m|}{2\pi^2} \sum_{\nu=0}^{\nu_i^\mathrm{max}}
(2-\delta_{\nu 0})  \frac{3}{2 R} \int_{\Lambda_\mathrm{IR}}^{p_F}  \frac{E_i-\mu_i}{p} \arctan(x_i) \mbox{d}p_z , \label{eq;osur}\\
B_{S,i} &=& \frac{d_i |q_iB_m|}{2\pi^2}
\sum_{\nu=0}^{\nu_i^\mathrm{max}} (2-\delta_{\nu 0}) \frac{3}{2 R}
\int_{m^*}^{\mu_i}\int_{\Lambda_\mathrm{IR}}^{p_F} \left[
\frac{E_i-\mu_i}{E_i^2} +\frac{m_i^*}{E_i
p}\arctan(x_i)\right]\frac{dm_i^*}{d\mu_i} \mbox{d}p_z \mbox{d}\mu_i
.\label{eq:Bsur}
\end{eqnarray} In the bulk limit $R\rightarrow \infty$ and for the light quark mass $m=a \mu$ with $a=g/\sqrt{6\pi^2}$, the bag function in Eq. (\ref{Bexp}) has the analytical expression as
\begin{eqnarray}
B_i^*(\mu_i)&=&  \frac{d_i |q_iB_m|}{4\pi^2}
\sum_{\nu=0}^{\nu_i^\mathrm{max}} (2-\delta_{\nu 0}) a^2\mu^2
\left[\frac{1}{2}\ln(\frac{1+k_z}{1-k_z})
+\frac{k^2-k_z^2}{a^2}\tanh^{-1}(\frac{1}{k_z})-\frac{k^2-k_z^2}{a^2}\tanh^{-1}(\frac{\sqrt{1-k_z^2}}{a
k_z})    \right. \nonumber\\
&& \left. \phantom{\frac{d_i |q_iB_m|}{4\pi^2}
\sum_{\nu=0}^{\nu_{max}} (2-\delta_{\nu 0}) a^2\mu^2 \left[ \right.
} +\frac{k^2-k_z^2}{a^2}\ln( \frac{k \sqrt{1-k_z^2}+a k_z}{k+k_z})
-(1-k_z^2)\ln[\frac{\sqrt{1-k_z^2}+a k_z}{\sqrt{1-k_z^2-a^2 k_z^2}}]
\right],
\end{eqnarray}
where we define a dimensionless momentum $k =\sqrt{1-a^2}$ and its
z-component in the magnetic field $k_z=\sqrt{1-a^2-2\nu
|q_iB_m|/\mu_i^2}$. If the magnetic field is so strong that all
quarks are lying on the lowest Landau level (LLL), the bag function
can be simplified as
\begin{eqnarray}
B_i^*(\mu_i)= -\frac{d_i |q_iB_m|}{4\pi^2} a^2 \mu^2 (1-a^2)
\ln(\frac{1+\sqrt{1-a^2}}{a}).
\end{eqnarray}

\subsection{The anisotropic surface tension}
In literature the surface tension was investigated with the two main
approaches. One is the multiple reflection expansion approximation
\cite{Shao:2006gz}. The other method is the geometric approach
\cite{Pinto:2012aq,Garcia:2013eaa}. The surface tension is
characteristic of the two phase. For example, the amount of the
surface tension between the liquid drop and the gas phase is
dominant to the raindrop formation. The surface tension between true
vacuum and perturbative phase was usually neglected in comparison to
the confinement bag constant. The early work is typically traced
back to the paper \cite{Berger:1986ps} by Berger et al. They
suggested that the surface tension parameter can be calculated from
the surface modification of the fermion density of states with
larger values of the surface tension could survive the early
Universe. The surface tension also depends strongly on the dynamical
effects and the skin thickness \cite{Parija:1993sq}. In contrast,
there is a suggestion that the surface tension for the interface
separating the quark and the hadron phase should be smaller to make
the mixed phase occur \cite{Maruyama:2007ey,Alford:2001zr}.
Recently, in order to employ the excess energy associated with
inhomogeneous configurations, the definition of a differential
surface tension from the bubble formation in the discrete case to
systems with continuous symmetry \cite{Endrodi:2021kur}. The special
value of the surface tension is poorly known. At least, it is
possible that there is a critical surface tension, above which the
structure of the mixed phase will become unstable
\cite{Voskresensky:2001jq}.

Once the thermodynamic potential is known, the longitudinal pressure
is obtained as
\begin{eqnarray} P^\parallel=-\Omega,
\end{eqnarray} As is mentioned that in the presence of the strong magnetic
field, the rotational symmetry breaking would be demonstrated not
only by the anisotropic pressure structure but also by the surface
tension for the finite volume matter. Therefore, the longitudinal
surface tension can be derived by the longitudinal pressure
\cite{Wen:2010zz}.
\begin{eqnarray}
 \sigma_i^\parallel=\frac{R}{3}(\Omega_{S,i}+B_{S,i}),
\end{eqnarray}
The electromagnetic contribution of Maxwell term $B_m^2/2$ is
omitted to the pressure due to that it has no influence on the
surface tension. The transverse pressure is usually defined as
$P^\perp=P^\parallel-M B_m$, where the magnetization susceptibility
$M$ can be derived by the relation $M=-\frac{\partial
\Omega}{\partial B}$. In the present work, we have the transverse
pressure from the $i$-flavor quarks
\begin{eqnarray}P_i^\perp &=& \frac{d_i |q_iB_m|^2}{2\pi^2} \sum_{\nu=1}^{\nu_i^\mathrm{max}}
\nu
  \left\{\int_{\Lambda_\mathrm{IR}}^{p_F}
    \frac{\rho_\text{MRE}\mbox{d}p_z}{E_i} \right. \nonumber \\
& & \left. \phantom{\frac{d_i |q_iB_m|^2}{2\pi^2}
\sum_{\nu=1}^{\nu_i^\mathrm{max}}\ nu [}-
\int_{m^*}^{\mu_i}\int_{\Lambda_\mathrm{IR}}^{p_F}
\Bigg[\frac{m}{E^3}+\frac{3}{R} \Big( \frac{1}{p^2 E_i}
-2\frac{\mu_i}{ E_i^4}+\frac{m}{p^3
E_i^3}(E_i^2+p^2)\arctan(\frac{m_i}{p})
\Big)\Bigg]\frac{dm_i^*}{d\mu_i} \mbox{d}p_z \mbox{d}\mu_i \right\}.
\end{eqnarray}

Similar to the expression of the longitudinal surface tension
$\sigma^\parallel$, the transverse surface tension contribution per
flavor can be divided into two parts as
\begin{eqnarray}\sigma_i^\perp=\frac{R}{3}(\Omega_{S,i}'+B_{S,i}'),
\end{eqnarray}
where the notation ${ O}'$ stands for a transformation of the
quantity $O$, namely, ${ O}'={ O} + {B_m} \frac{\partial
{O}}{\partial B_m}$. Applying the operation on both the free term
$\Omega$ and the bag function $B^*$, one can find the following
expressions
\begin{eqnarray}\Omega_{S,i}' &=& \frac{d_i |q_iB_m|^2}{2\pi^2} \sum_{\nu=1}^{\nu_i^\mathrm{max}}
\nu  \int_{\Lambda_\mathrm{IR}}^{p_F}
    \frac{3}{Rp} \arctan(x_i)\frac{\mbox{d}p_z}{E_i}.\\
B_{S,i}' &=& \frac{d_i |q_iB_m|^2}{2\pi^2}
\sum_{\nu=0}^{\nu_i^\mathrm{max}} \nu \frac{3}{R}
\int_{m^*}^{\mu_i}\int_{\Lambda_\mathrm{IR}}^{p_F} \left[
\frac{1}{p^2 E_i} -2\frac{\mu_i}{ E_i^4}+\frac{m}{p^3
E_i^3}(E_i^2+p^2)\arctan(\frac{m_i}{p}) \right]\frac{dm_i^*}{d\mu_i}
\mbox{d}p_z \mbox{d}\mu_i.
\end{eqnarray}
It can be well understood that the anisotropy of dense quark matter
may stem from the magnetization along the field direction resulted
by two reasons, namely, the arrangement of free particles and the
medium effect.


\section{Numerical result and conclusion}\label{sec:result}
In the present paper we take the current mass $m_{0}$=5.6 MeV and
$\mu_u=\mu_d=\mu$ for isospin-symmetric quarks and the bag constant
$B_0=(145$MeV)$^4$ for bulk case. The main advantage of the
quasiparticle model is the combination of the medium effect into
both the effective quark mass and the bag function $B^*$. For the
intensity of the field $eB_m<\mu^2$, the validity of the spherical
geometry could be approximately available to some extent.

\begin{figure}[H]
\centering
        \includegraphics[width=8cm]{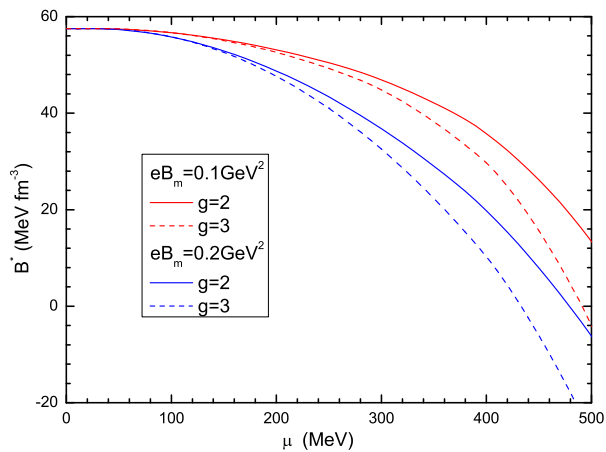}
\caption{The effective bag function $B^*(\mu)$ at the coupling
constant is shown as a function of the chemical
potential.}\label{fig1}
\end{figure}

\begin{figure}[H]
    \centering
    \includegraphics[width=8cm]{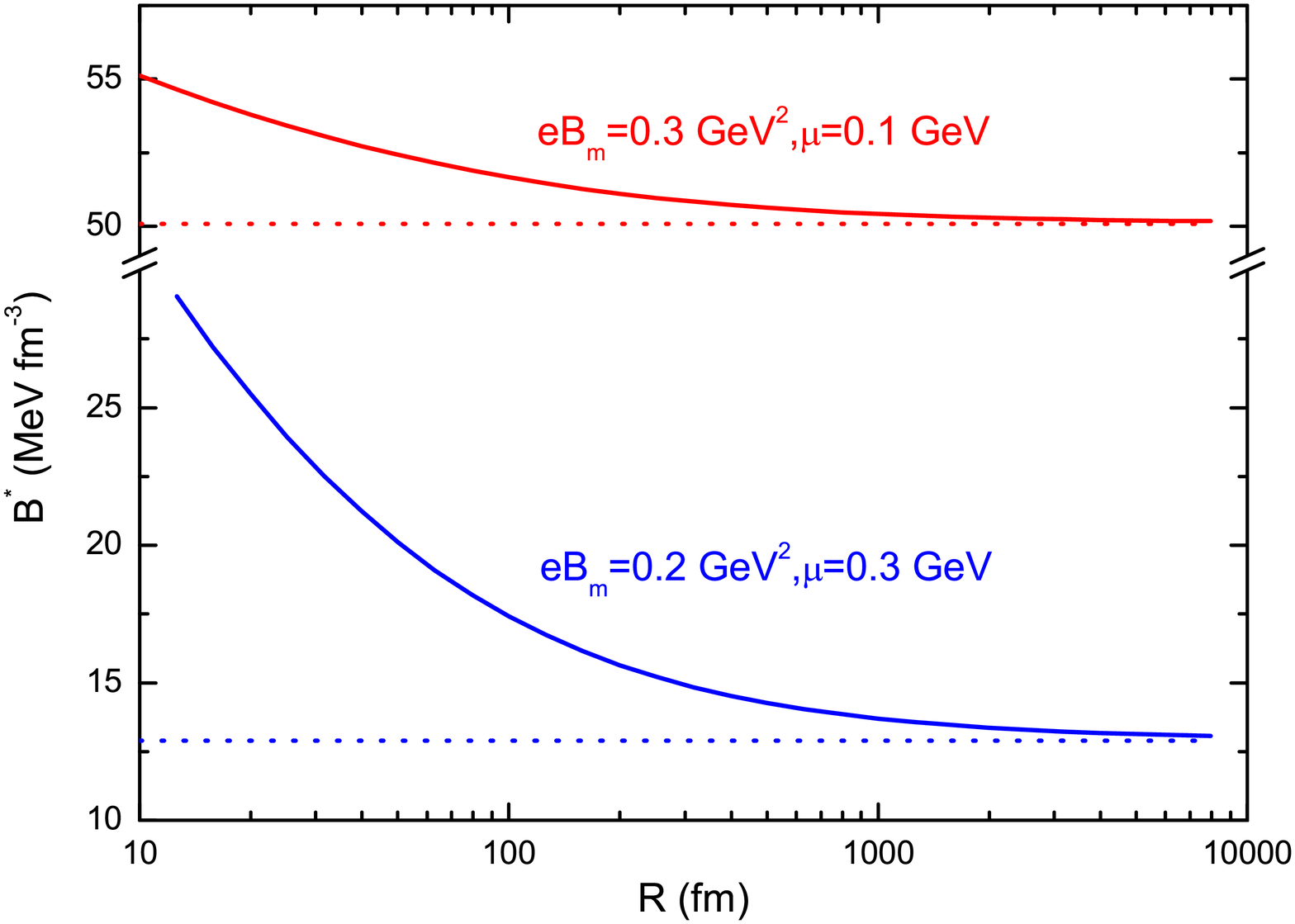}
    \caption{\label{fig2}The effective bag function $B^*(\mu)$ at
the coupling constant is shown as a function of the radius of the
spherical system.}
\end{figure}

The chemical potential dependent bag function plays an important
role in the description of the deconfinement transition. The bag
function would play as a function of the chemical potential and the
finite size. The magnetic field effect and the coupling constant are
considered separately. In Fig.~\ref{fig1}, the bag function
decreases with the increase of the chemical potential, which
indicates a signal of the deconfinement transition. As the coupling
constant becomes larger and/or the magnetic field becomes stronger,
the decreasing behavior of the bag function would happen at a
smaller chemical potential, which indicates a critical chemical
potential is similar to the temperature behavior characterized by
the inverse magnetic catalysis effect \cite{Preis:2010cq}. In
Fig.~\ref{fig2}, the effect of the finite size volume is shown on
the bag function. The two horizontal dotted lines are the
corresponding bag functions for bulk strange quark matter. The bag
function increases as the spherical size decreases. As the spherical
radius $R$ approach the infinite value, the bag function is
gradually close to the constant $B_0$ for the bulk matter, which
indicates the disappearance of the finite size effect.

\begin{figure}[H]
    \centering
    \includegraphics[width=8cm]{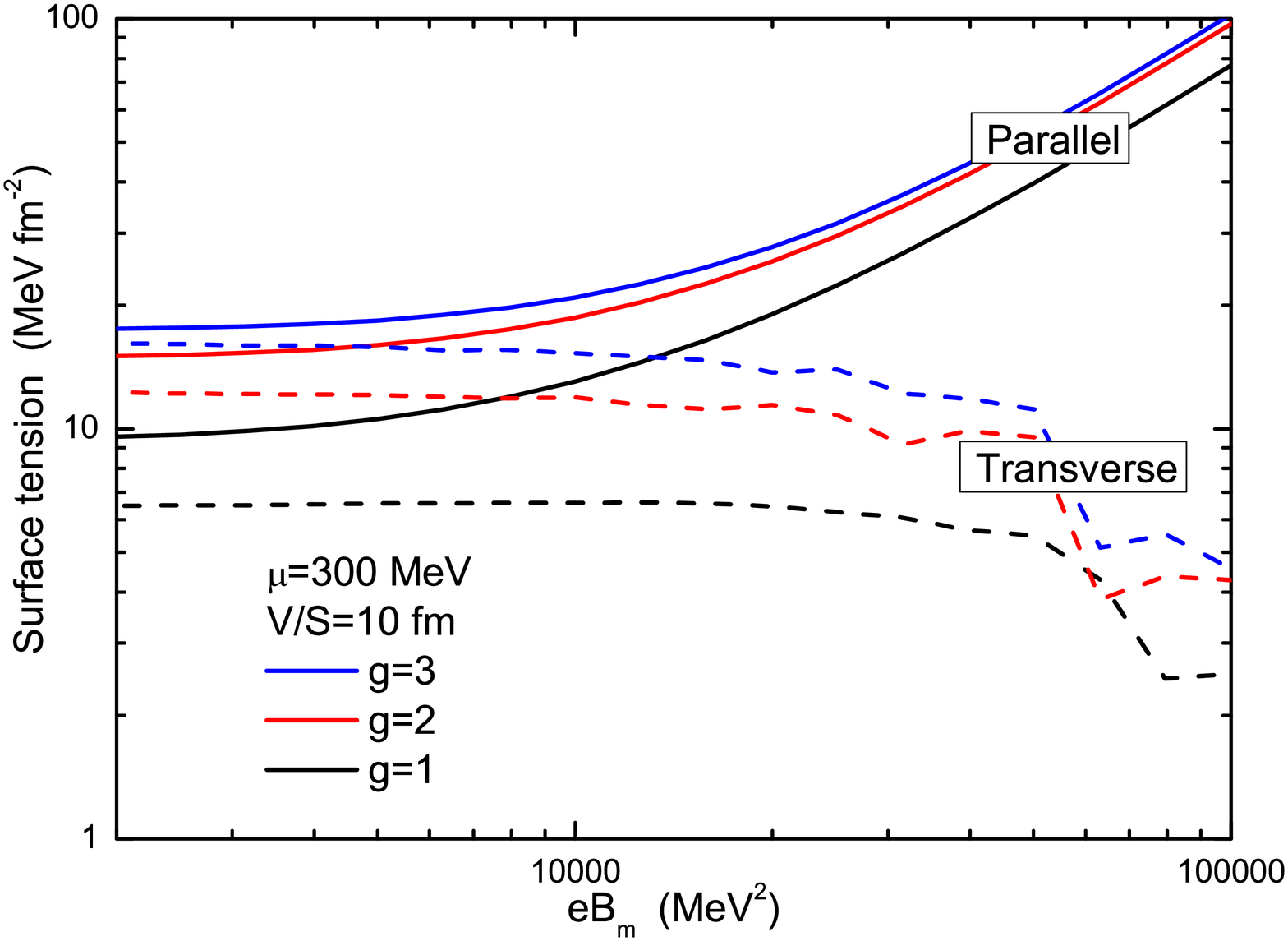}
    \caption{\label{fig3}The parallel and transverse surface tensions with $V/S=10$~fm and $\mu=300$~MeV at vanishing temperature are shown as function of the magnetic field strength.}
\end{figure}

At the larger chemical potential, a direct transition from the
vacuum to quark matter happens possibly depending on the surface
tension of bubble quark matter \cite{Fraga:2018cvr}. The surface
tension is plotted as a function of the magnetic field in Fig.
\ref{fig3}. The chemical potential and the finite size are adopted
as $\mu=300$ MeV and $V/S=10$ fm. The different coupling constants
$g=1$, 2, and 3 are marked by the black, the red, and the blue
curves from bottom to top. The longitudinal and transverse surface
tensions are denoted by the solid and dashed curves respectively. It
can be found that the longitudinal component increases with the
magnetic field $eB_m$ and coupling constant $g$. In particular, the
longitudinal surface tension is proportional to the magnetic field
in the strong field limit. It can be simply understood that the
magnetic field only has contribution to the coefficient in front of
the integral of Eqs. (\ref{eq;osur}) (\ref{eq:Bsur}). On the
contrary, the transverse surface tension would decrease as the
increasing magnetic field. Furthermore, the transverse component
feels the Landau level effect more sensitively, which results in a
oscillation behavior. At weak magnetic field, the value of the
surface tension is in agreement with the result estimated as the
order as $(70~$MeV)$^3$ \cite{Farhi:1984qu}. As the magnetic field
becomes much stronger, the anisotropy structure would be enhanced
greatly.

\begin{figure}[H]
    \centering
    \includegraphics[width=8cm]{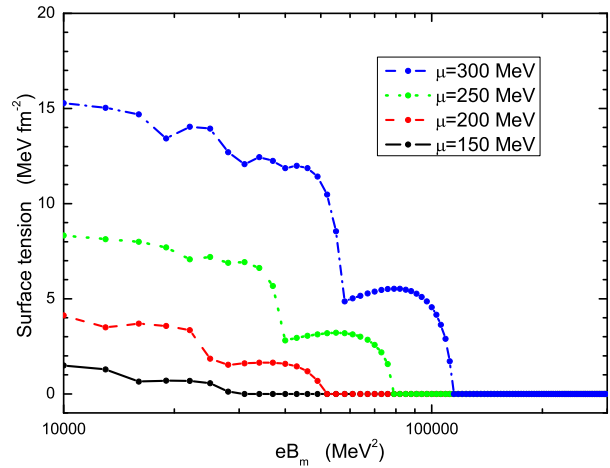}
    \caption{\label{fig4}The behavior of the transverse surface tension $\sigma^\perp$ is shown as a function of the magnetic field at four different chemical potentials with $V/S=10$ fm.}
\end{figure}
In Fig. \ref{fig4}, the transverse surface tension is plotted as
function of the magnetic field at four different chemical potentials
$\mu=150$, 200, 250, 300 MeV and the fixed size $V/S=10$ fm. The
oscillation behavior is shown clearly in the weaker magnetic field.
The more quarks located in high Landau levels have a finite
contribution to the motion perpendicular to the magnetic field. At
the much stronger magnetic field, the transverse surface tension
would drop down and vanish in the end, which can be understood that
all quarks located in the LLL have no contribution to the transverse
motion. It is suggested that the vanishing of the transverse surface
tension is a signal of all charged particle occupied in the LLL.
Moreover, the threshold value of the magnetic field for the
vanishing of $\sigma^\perp$ would increase with the increase of the
chemical potential.

\begin{figure}[H]
    \centering
    \includegraphics[width=8cm]{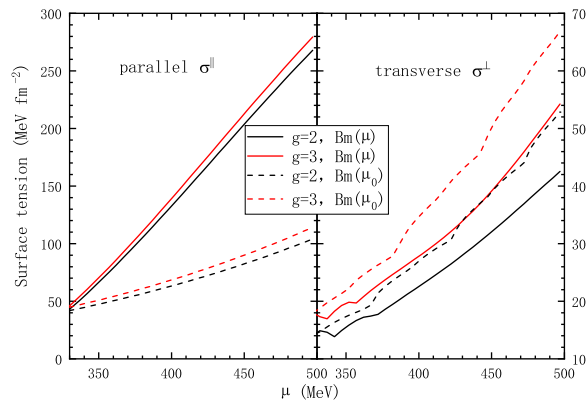}
    \caption{\label{fig5}The parallel and transverse surface tensions are shown as functions of the chemical potential in a uniform magnetic field $B_\mathrm{m}(\mu_0)$ and a nonuniform $B_\mathrm{m}(\mu)$ \cite{Dexheimer:2016yqu} }
\end{figure}

To mimic a realistic magnetic field in compact stars, the magnetic
field is recently suggested to increase polynomial instead of
exponential as the chemical potential \cite{Dexheimer:2016yqu}. In
Fig. \ref{fig5}, the anisotropic surface tension is investigated as
a function of the chemical potential. The nonuniform magnetic field
$B_\mathrm{m}(\mu)$ and the fixed strength $B_\mathrm{m}(\mu_0)$ are
marked by the solid lines and the dashed lines, respectively. The
initial point $\mu_0$ is associated to the surface chemical
potential $\mu_0$ of compact stars
\cite{Dexheimer:2016yqu,Peterson:2023bmr}. The parallel surface
tension $\sigma^\parallel$ on left panel and transverse one
$\sigma^\perp$ on right panel are calculated at different couplings
$g=2$ and 3. The transverse surface tension is apparently smaller
than the parallel one in the whole range of the chemical potential,
which reflects the anisotropic structure. It should be emphasized
that the transverse surface tension would decrease as the increasing
the magnetic field at high densities, which is indicated in Fig.
\ref{fig4}. However, the surface tension is enlarged by the
increasing chemical potential. Therefore, the transverse surface
tension is shown as an increasing function of the chemical
potential. The tiny oscillation behavior occurs at low chemical
potential due to the Landau level transition. The transverse surface
tension under the magnetic field $B_\mathrm{m}(\mu)$ is smaller than
that of the case $B_\mathrm{m}(\mu_0)$. On the contrary, the
parallel tension is larger and grows more rapidly under the the
magnetic field profile $B_\mathrm{m}(\mu)$.
\section{Summary}

In this paper, the thermodynamics of magnetized quark matter in the
finite size has been obtained in the quasiparticle model. The dense
medium effect is included through the effective bag function, which
depends on the chemical potential and the magnetic field. The bag
function plays as an appropriately chosen vacuum energy constant
ensuring thermodynamic consistency. On the other hand, it provides a
measure for nonperturbative physics which cannot be described by the
effective masses. Its variation would be helpful to investigate the
deconfinement transition. As expected, it has been numerically shown
that the bag function is a decreasing function of the chemical
potential. The drop down behavior would be strengthened by the
increase of both the magnetic field and the coupling interaction
constant. It was verified that the bag function $B^*$ would
gradually approach the bulk limit $B_0$ as the size becomes
infinitely large. We have employed the extended quasiparticle model
to the surface tension. It is found that the medium effect
represented by the bag function would lead to an additional term in
the surface tension. The anisotropy of the longitudinal and
transverse surface tensions is enhanced by an increase of the
magnetic field. The longitudinal surface tension is an increasing
function of the magnetic field. But the transverse component would
decrease and drop down to zero at extreme strong magnetic fields.
The vanishing of transverse surface tension coincides with the
condensation of all quarks in LLL. Finally, the new expression of
the surface tension modified by the medium effect would be useful to
generalize the current investigation on the bubble formation in the
QCD transition. Last but no least, the infrared cutoff for avoiding
the negative density of state is relevant with the confinement
phenomenon. The more reasonable confinement mechanism should be
produced by the self-consistent combination of the infrared cutoff
and the bag constant in future work.

\acknowledgments{ The authors would like to thank support from the
National Natural Science Foundation of China under the Grants No.
11875181, No. 11705163, and No. 12147215. This work was also
sponsored by the Fund for Shanxi "1331 Project" Key Subjects
Construction. }

\end{document}